\documentclass[aps,prb,amsmath,twocolumn,amssymb,floatfixng,showpacs,
superscriptaddress,footinbib]{revtex4-1}
\pdfoutput=1
\usepackage[dvips]{graphics}
\usepackage{bm}
\usepackage{epsfig}
\usepackage{enumerate}
\usepackage{subfigure}
\usepackage{color}
\usepackage{bbold}
\usepackage{graphicx}
\usepackage{hyperref}
\hypersetup{
    pdfnewwindow=true,      
    colorlinks=true,       
    linkcolor=magenta,          
    citecolor=blue,        
    filecolor=blue,      
    urlcolor=blue        
}

\newcommand{\ie}{{\it i.e., }}

\newcommand{\la}{{\langle }}
\newcommand{\ra}{{\rangle }}
\newcommand{\eps}{{\epsilon }}
\newcommand{\vareps}{{\varepsilon }}

\begin{document} 
\title{Magnetotransport properties of ${\rm 8-Pmmn}$ borophene: effects of Hall field and strain} 

\author{SK Firoz Islam}
\email{firoz@iopb.res.in}
\affiliation{Institute of Physics, Sachivalaya Marg, Bhubaneswar-751005, India}

\begin{abstract}
The polymorph of ${\rm 8-Pmmn}$ borophene is an anisotropic Dirac material with tilted
Dirac cones at two valleys. The tilting of the Dirac cones at two valleys are in opposite
directions, which manifests itself via the valley dependent Landau levels in presence of an
in-plane electric field (Hall field). The valley dependent Landau levels cause valley
polarized magnetotransport properties in presence of the Hall field, which is in contrast
to the monolayer graphene with isotropic non-tilted Dirac cones. The longitudinal conductivity and Hall conductivity are evaluated by using
linear response theory in low temperature regime. An analytical approximate form of the
longitudinal conductivity is also obtained. It is observed that the tilting of the Dirac cones
amplifies the frequency of the longitudinal conductivity oscillation (Shubnikov-de Haas).
On the other hand, the Hall conductivity exhibits graphene-like plateaus except the appearance
of valley dependent steps which are purely attributed to the Hall field induced lifting of the
valley degeneracy in the Landau levels. Finally we look into the different cases when the Hall field
is applied to the strained borophene and find that valley dependency is fully dominated by strain
rather than Hall field. Another noticeable point is that if the real magnetic field is replaced by the strain induced 
pseudo magnetic field then the electric field looses its ability to cause valley polarized transport.
\end{abstract}

\maketitle
\section{Introduction}
The discovery of the atomically thin two dimensional (2D) material-graphene\cite{RevModPhys.81.109,sarma2011electronic},
has received much attention in the last decade because of its unique physical properties as well as
possible future applications. The electronic properties of the graphene are governed by its massless
linear band dispersion rather than usual parabolic. Moreover, another intriguing 
feature of the graphene is that the bulk consists two inequivalent valleys ({$K$ and $K^{\prime}$) in the
first Brillouin zone  of its band structure, which are the key ingredient for the newly emerged field -Valleytronics
\cite{schaibley2016valleytronics,PhysRevB.96.245410,PhysRevLett.116.016802}, just like Spintronics
\cite{RevModPhys.76.323,RevModPhys.87.1213,RevModPhys.89.011001} based on spin.
Apart from the graphene, similar materials like silicene\cite{lalmi2010epitaxial,
PhysRevLett.108.155501,PhysRevB.85.075423,PhysRevLett.107.076802},
transition-metal dichalcogenides\cite{PhysRevLett.108.196802,PhysRevLett.105.136805}, exhibiting linear
dispersion with strong spin-orbit coupling, have also been extensively considered from the theoretical
as well as experimental fronts. The polymorph of borophene  which exhibits anisotropic 
tilted Dirac cones in its band structure (named as ${\rm 8-Pmmn}$) is the latest member to the
family of Dirac materials \cite{PhysRevLett.112.085502} after the experimental
realization\cite{PhysRevLett.118.096401} of it. A detailed {\it ab-initio}
properties\cite{PhysRevB.93.241405} of this material was also addressed recently.
Similar to the strained graphene\cite{PhysRevLett.103.046801},
a pseudo magnetic field has also been predicted in ${\rm 8-Pmmn}$ borophene\cite{PhysRevB.94.165403}
under the influence of strain.
Very recently, several theoretical investigations on optical properties like anisotropic 
plasmons\cite{PhysRevB.96.035410}, effects of particle-hole symmetry breaking in optical
conductivity\cite{PhysRevB.96.155418} and Drude weight have been reported.

The magnetotresistivity measurement of a 2D electronic system is one of the most appreciated method 
to probe the system. The application of a perpendicular uniform magnetic field to the 2D
electronic systems quantizes the electronic energy spectrum \ie forms Landau levels (LLs). 
The LLs can be realized by oscillatory longitudinal conductivity with inverse magnetic field known as
Shubnikov-de Haas (SdH) oscillations\cite{feng2005introduction,imry1997introduction}. On the other hand,
the off-diagonal terms in conductivity tensor becomes quantized due to the incomplete cyclotron orbits
along the opposite transverse edges of the system\cite{feng2005introduction,imry1997introduction}.
The quantum Hall conductivity in graphene \cite{PhysRevLett.95.146801,PhysRevB.65.245420,PhysRevB.86.115432}
is $\sigma_{xy}= 4(n+1/2)e^2/h$ with $n=0,1,2,3..$, which is in contrast to usual 2D electron
gas where $\sigma_{xy}=2(n+1)e^2/h$. Note that `e' and `h' are the electronic charge and the Planck constant, respectively. 
Apart from the graphene, the magnetoconductivity has been extensively studied in 
silicene\cite{PhysRevB.90.235423,tahir2013valley}, topological insulators\cite{tahir2012quantum,
PhysRevB.92.045416,buttner2011single,zhang2015edge},phosphorene\cite{zhou2015landau,PhysRevB.92.165409,PhysRevB.96.085434},
stanene\cite{doi:10.1002/pssb.201552341} and molybdenum disulfide\cite{tahir2016quantum,Firoz} etc.
Apart from the modulation induced Weiss oscillations in ${\rm 8-Pmmn}$
borophene\cite{Firoz_weiss}, several theoretical 
investigations\cite{PhysRevB.91.195413,morinari2009possible,kobayashi2008hall}
of magnetotransport properties in 2D Dirac materials with tilted Dirac cones have been also carried out.
However, so far no attempt has been made in such material with tilted Dirac cones to modulate valley degree of
freedom in the integer quantum Hall effect and the longitudinal conductivity by applying an
in-plane electric field (Hall field) in presence of randomly scattered charge impurities. In this work,
we rectify this anomaly and try to obtain valley dependent magnetoconductivity in presence of a Hall field.

In this work, we investigate the quantum magnetotransport properties in presence of a Hall field in low
temperature regime by using the linear response theory. One of the key issues of the valleytronics (spintronics)
is how to control or modulate the two valleys (spin) independently by means of external parameters.
We aim to modulate the valley dependency of the magnetoconductivity by applying an in-plane electric field. 
We should mention here that such valley dependent transport can be found in other 2D Dirac
materials too\cite{yarmohammdai1,HOI2017203,HOI2018340,yar4}. However, in those materials the presence of 
a strong spin-orbit interaction term removes the spin/valley degeneracy in its band structure.
Assuming the elastic or quasi-elastic scattering of electron by charge impurities, scattered
randomly in the system, we calculate the longitudinal and the Hall conductivity. 
The longitudinal conductivity shows SdH oscillations with the inverse magnetic field.
The frequency of the SdH oscillations is amplified by the tilting of the Dirac cones. We also notice
that the quantum Hall conductivity exhibits Hall plateaus of the form of $\sigma_{xy} =2(n+1/2)e^2/h$,
exactly similar to the case of graphene in each valley. However, a valley separation is visible at the Hall
steps and the SdH oscillations peaks due to the presence of the Hall field. This is in contrast to
the non-tilted isotropic Dirac material like graphene\cite{PhysRevB.83.075427}, where magnetotransport
properties are not sensitive to the valley index in the presence of a Hall field. 

The paper is organized as follows. In Sec.~\ref{sec2}, we introduce the low energy effective Hamiltonian and
discuss the lifting of the valley degeneracy in LLs in presence of a Hall field.
The Sec.~\ref{sec3} is devoted to calculate different components of the magnetoconductivity
tensor and analyze the results. Finally, we summarize and conclude in Sec.~\ref{sec4}.

\section{Model Hamiltonian and Landau level formation}\label{sec2}
In this section, we derive LLs and corresponding eigen states. We start with the low-energy single-particle 
effective model Hamiltonian for the tilted anisotropic Dirac cones as\cite{PhysRevB.94.165403,PhysRevB.96.035410}
\begin{equation}
 H=\xi(v_xp_x\sigma_x+v_yp_y\sigma_y+v_tp_y\mathbb{1}),
\end{equation}
where $\xi=+(-)$ denotes the valley $K(K')$, three velocities are given by $\{v_x,v_y,v_t\}=\{0.86,0.69,0.32\}$
in units of $v_0=10^6$ m/sec. Also, ${\bf\sigma\equiv(\sigma_x,\sigma_y)}$ are the pseudo Pauli matrices and $\mathbb{1}$
is identity matrix. Note that unlike {\it non-tilted isotropic} Dirac cones in graphene,
the  velocities along $x$ and $y$ direction are not identical. The above Hamiltonian
can be diagonalized to obtain the energy dispersion as
 \begin{equation}\label{band2}
  \vareps_{\lambda,k}^{\xi}=\xi\hbar v_tk_y+\lambda\hbar\sqrt{v_x^2k_x^2+v_y^2k_y^2},
 \end{equation}
where $\lambda=\pm$ denotes the band index and ${\bf k}=\{k_x,k_y\}$ is the $2$D momentum vector.
\begin{figure}[!thpb]
\centering
\includegraphics[height=5cm,width=0.90 \linewidth]{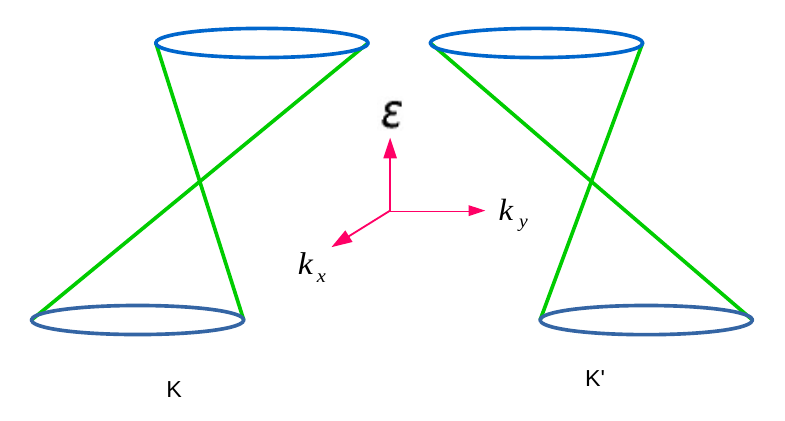}
\caption{(Color online) A schematic sketch of the energy band dispersion in k-space, representing Eq.~(\ref{band2})
in valley K and K$^{\prime}$.}
\label{band}
\end{figure}
This energy dispersion is shown schematically in Fig.~(\ref{band}), which is tilted along $k_y$-direction
due to the presence of $v_t$ term. However, the tilting of the two Dirac cones at the two valleys is
in opposite directions. Note that the tilting of the Dirac cones breaks the particle-hole symmetry
in the $8-Pmmn$ borophene.
\subsubsection{Inclusion of crossed electric and magnetic field}
To include the effects of perpendicular magnetic field (${\bf B}=B\hat{z}$) in the low 
energy single-electron effective Hamiltonian of borophene, lying in the $x$-$y$ plane,
we use the Landau-Peierls substitution, ${\bf p}\rightarrow {\bf p+eA}$,  as
\begin{equation}
\mathcal{H}=\xi[v_xp_x\sigma_x+v_y(p_y+eBx)\sigma_y+v_t(p_y+eBx)\mathbb{1}]
\end{equation}
under the Landau gauge ${\bf A}=(0,xB,0)$. Here, $A$ represents the magnetic vector potential. The effect of an in-plane
uniform real electric field ($E_r$) can be included by adding a potential energy $U=eE_{r}x\mathbb{1}$ to the
Hamiltonian as
\begin{equation}\label{hamil}
\mathcal{H}_{e}=eE_{r}x\mathbb{1}+\xi[v_xp_x\sigma_x+v_y(p_y+eBx)\sigma_y+v_t(p_y+eBx)\mathbb{1}],
\end{equation}
 The Hamiltonian is translationally invariant along the $y$-direction as $[\mathcal{H}_{e},p_y]=0$, which allows
 the electron to be governed by the wave function $\Psi(x,y)\sim e^{ik_yy}\phi(x)$. Using this fact,
 the eigen value problem reduces to
\begin{equation}
 \mathcal{H}_{0}\phi(x)=\vareps\phi(x),
\end{equation}
where
\begin{equation}\label{hamil}
\mathcal{H}_{0}=\frac{\hbar v_{e}^\xi}{l_c}X\mathbb{1}+\xi\left\{\frac{\hbar v_c}{l_c}
\left[\sqrt{\frac{v_x}{v_y}}\sigma_xP+\sqrt{\frac{v_y}{v_x}}\sigma_y X\right]\right\}, 
\end{equation}
and 
\begin{equation}
\vareps=E+\hbar v_{e}k_y,
\end{equation}
where $v_{e}^{\xi}=v_{e}+\xi v_t$ with $v_{e}=E_{r}/B$, the magnetic length $l_c=\sqrt{\hbar/eB}$, 
the dimensionless x-component of momentum operator $P=-i\partial/\partial(x/l_c)$,
position operator $X=(x+x_0)/l_c$ with the center of cyclotron orbit is at $x=-x_0=-k_yl_c^2$ and $v_c=\sqrt{v_xv_y}$.
Apart from the velocity anisotropy inside the third bracket in Eq.~(\ref{hamil}), the above Hamiltonian is very
much identical to the case of monolayer graphene under crossed electric and magnetic field\cite{lukose2007novel}.
The first term acts as a pseudo in-plane effective electric field ($E_{eff}=\hbar v_{e}^{\xi}/(el_{c}^2)$).
Now Eq.~(\ref{hamil}) can be re-written as
\begin{equation}
 H_{\xi}=e\frac{E_{eff}}{\sqrt{2}}(a+a^{\dagger})\mathbb{1}+
 \xi\hbar \omega_c\left[\begin{array}[c]{c c}0 & -ia\\ia^{\dagger} &0\end{array}\right],
\end{equation}
where $\omega_c (=v_c/l_c)$ is the cyclotron frequency and ladder operators are defined as:
$a=(\tilde{X}+i\tilde{P})/\sqrt{2}$ and $a^{\dagger}=(\tilde{X}-i\tilde{P})/\sqrt{2}$.
Here, $\tilde{X}=\sqrt{\frac{v_y}{v_x}}X$ and $\tilde{P}=\sqrt{\frac{v_x}{v_y}}P$, satisfying the 
commutator relation $[\tilde{X},\tilde{P}]=i$. In absence of $E_{eff}$, the above Hamiltonian
can be diagonalized to obtain graphene-like LLs (for $n\ge 0$) 
\begin{equation}
 \vareps_{\zeta}=\lambda\hbar\omega_c\sqrt{2n},
\end{equation}
with $\zeta=\{n,\xi,k_y\}$ and eigenfunctions as
\begin{equation}
\Psi_{\zeta}({\bf r})=\frac{e^{ik_yy}}{\sqrt{2L_y}}\left[\begin{array}[c]{c}
                                  \xi\lambda\phi_{n-1}(X)\\
                                  i\phi_{n}(X)
                                  \end{array}\right],
\end{equation}
where $\phi_{n}(X)$ is the well known simple harmonic oscillator wave functions. The ground state ($n=0$)
wave function is
\begin{equation}
\Psi_{0,k_y}({\bf r})=\frac{e^{ik_yy}}{\sqrt{2L_y}}\left[\begin{array}[c]{c}
                                  0\\
                                  i\phi_{0}(X)
                                  \end{array}\right],
\end{equation}
In presence of $E_{eff}$, direct diagonalization of the above Hamiltonian is quite unwieldy. However,
a standard approach to solve this problem exactly was given by Lukose et al., in Ref.~[\onlinecite{lukose2007novel}].
Following this Ref.~[\onlinecite{lukose2007novel}], the above Hamiltonian can now be transformed into a frame,
moving along the y-direction with velocity $v_{e}^{\xi}$, such that the transformed electric field vanishes
and the magnetic field rescales itself as $B'=B\sqrt{1-\beta_{\xi}^2}$, where $\beta_\xi=v_{e}^{\xi}/\sqrt{v_xv_y}$.
Unlike the graphene, the noticeable point here is that the valley index is now intrinsically associated with 
the velocity of the moving frame, as well as in the renormalization of the transformed magnetic field in
moving frame. In the moving frame, LLs can be easily expressed as 
\begin{equation}
\tilde{\vareps}_{n,\xi,\tilde{k}_y}= \lambda\hbar\omega_c\sqrt{2n}(1-\beta_\xi^2)^{1/4}.
\end{equation}
However, to work in the rest frame, LLs must be brought back to this frame by using Lorentz boost back transformation
which gives LLs in rest frame as
\begin{equation}\label{ll}
 E_{\zeta}=\lambda\hbar\omega_c\sqrt{2n}(1-\beta_\xi^2)^{3/4}-\hbar v_e k_y
\end{equation}
and the eigen states are\cite{lukose2007novel,PhysRevB.92.035306}
\begin{eqnarray}\label{wave}
\Psi_{\zeta}({\bf r})&=&\frac{e^{ik_yy}}{\sqrt{2L_y\gamma_\xi}}\Big[\left(\begin{array}[c]{c}
                                  \cosh(\theta_\xi/2)\\
                                  -i\sinh(\theta_\xi/2)
                                  \end{array}\right)\lambda \phi_{n}(X')\nonumber\\
                                  &&-i\xi\left(\begin{array}[c]{c}
                                  i\sinh(\theta_\xi/2)\\
                                  \cosh(\theta_\xi/2)
                                  \end{array}\right)\phi_{n-1}(X')\Big]
\end{eqnarray}
with $\tanh\theta_\xi=\beta_\xi$ with $\gamma_\xi=1/\sqrt{1-\beta_\xi^2}$. 
We have also used the fact that the wave function in the rest frame differs from moving frame
by an imaginary phase factor $\exp[-(\theta_{\xi}/2)\sigma_y]$-hyperbolic rotation matrix, which is expressed as 
\begin{equation}
 e^{-\left(\theta_\xi/2\right)\sigma_y}=\left[\begin{array}[c]{c c}
                                  \cosh(\theta_\xi/2)&i\sinh(\theta_\xi/2)\\
                                  -i\sinh(\theta_\xi/2)&\cosh(\theta_\xi/2)
                                  \end{array}\right].
\end{equation}
On the other hand, the argument of the wave functions becomes 
\begin{equation}
 X'=\frac{(1-\beta_\xi^2)^{1/4}}{l_c}\left[x+k_yl_c^2+\lambda\frac{\sqrt{2n}l_c\beta_\xi}{(1-\beta_\xi^2)^{1/4}}\right]
\end{equation}
after using the Lorentz back transformation of momentum. The ground state ($n=0$ level) wave function is
\begin{eqnarray}\label{wave}
\Psi_{\{0,k_y,\xi\}}({\bf r})&=&\frac{e^{ik_yy}}{\sqrt{2L_y\gamma_\xi}}\Big[\lambda\left(\begin{array}[c]{c}
                                  \cosh(\theta_\xi/2)\\
                                  -i\sinh(\theta_\xi/2)
                                  \end{array}\right)\phi_{0}(X')\Big]\nonumber\\
\end{eqnarray}
with energy $E_{0,k_y}=-\hbar v_e k_y$. The LLs, derived in Eq.~(\ref{ll}), is sensitive to the valley index.
On the other hand, LLs in graphene\cite{lukose2007novel} under the influence of a Hall field is independent
of the valley index.
In graphene, under the suitable strength of the Hall field (for $\beta=E_r/(v_FB)=1$)
LLs get collapsed in both valleys. Whereas, similar situation can
appear in borophene but two valleys require different Hall fields $E_{r}^{c}=B(v_c\mp v_t)$.

The key point of this work is the lifting valley degeneracy in the LLs of a 2D Dirac materials, exhibiting
tilted Dirac cones, by applying an in-plane electric field. This was first pointed out by Goerbig's group\cite{goerbigEPL}
in an organic compound $\alpha-({\rm BEDT-TTF})_2{\rm I}_3$ having quite similar band structure.
Here, we exploit this issue in the magnetotransport properties of borophene.

\subsubsection{Density of states}
Before we proceed to magnetoconductivity, we shall examine the behavior of density of systes (DOS)
under the influence of an in-plane electric field. Because of the discrete energy levels i.e., LLs,
the DOS can be expressed as the sum of a series of delta function as 
\begin{equation}\label{delta}
D(E)=\frac{g_s}{\Omega}\sum_{\zeta}\delta(E-E_{\zeta})
\end{equation}
with $g_s$ is the spin degeneracy and the area of the system is denoted by $\Omega=L_x\times L_y$.
However, to plot the DOS in each valley, we assume impurity induced Gaussian broadening of the 
LLs and subsequently the Eq.~(\ref{delta}) simplifies to
\begin{equation}\label{dos}
 D(E)= D_{0}\sum_{n}\exp\left[-\frac{(E-E_n)^2}{2\Gamma_0^2}\right]
\end{equation}
where 
\begin{equation}
D_0=\frac{g_s}{2\pi l_c^2}\frac{1}{\Gamma_0\sqrt{2\pi}}.
\end{equation}
\begin{figure*}
\subfigure[]
{\includegraphics[width=.49\textwidth,height=6cm]{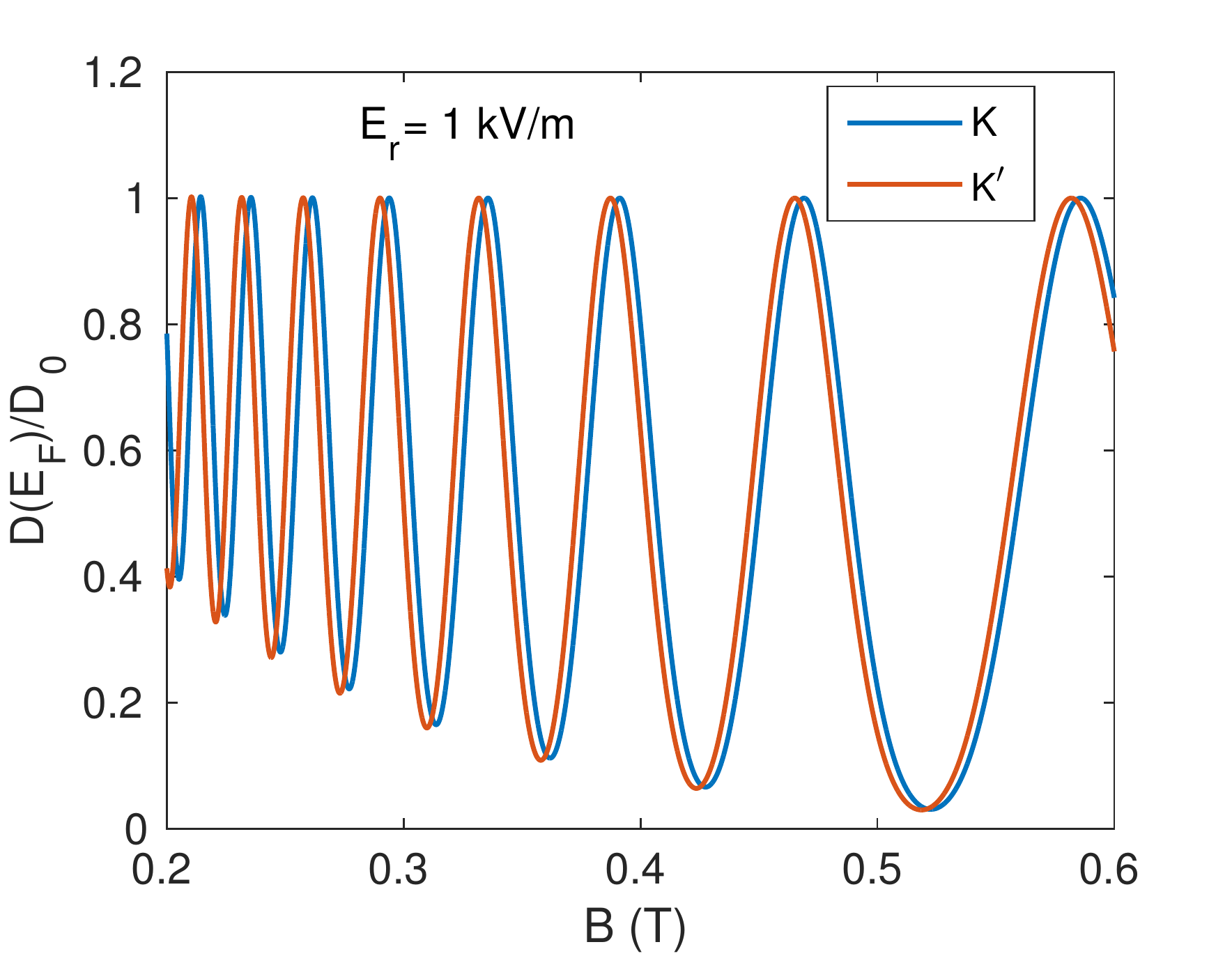}}
\subfigure[]
{\includegraphics[width=.49\textwidth,height=6cm]{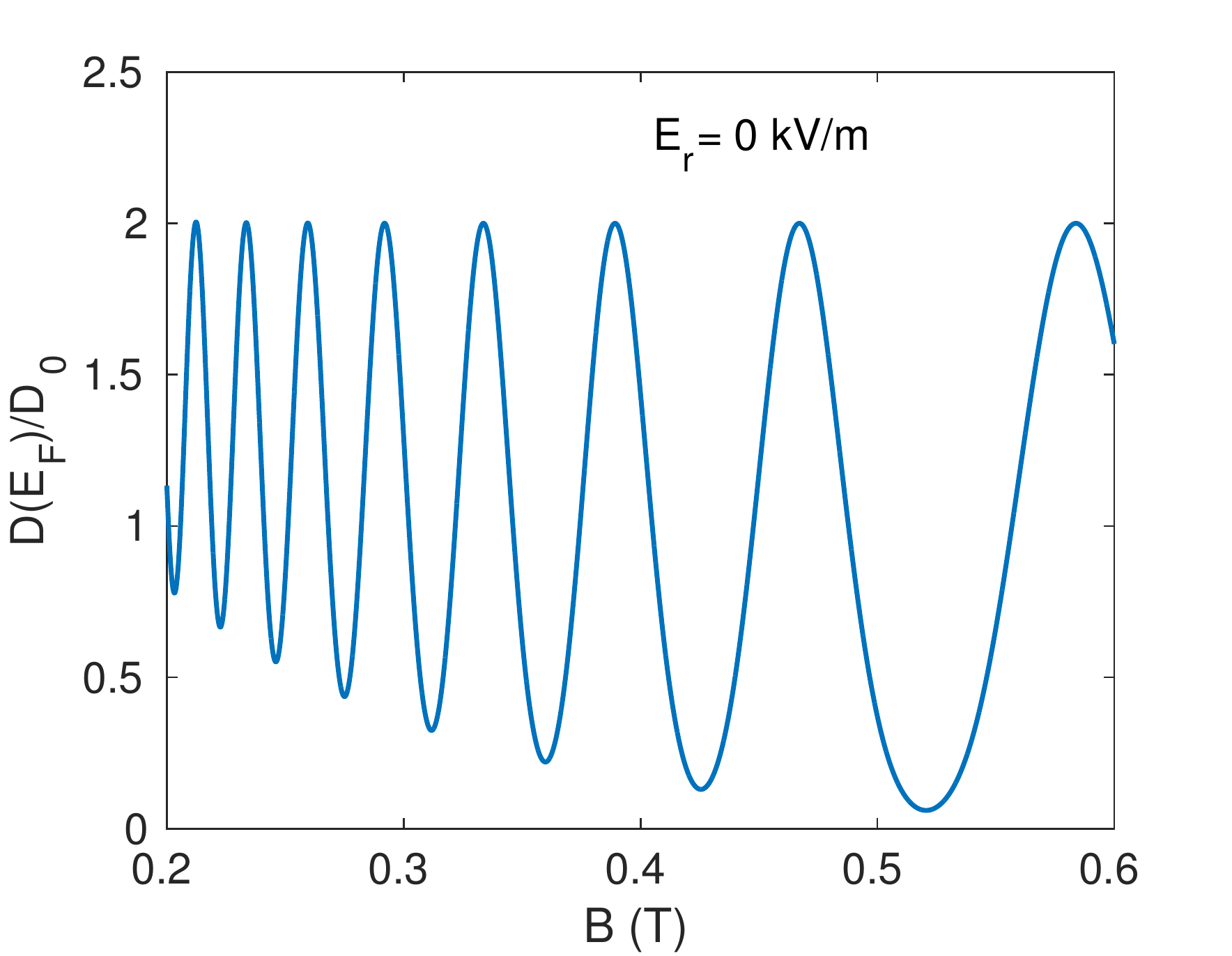}}
\caption{DOS versus magnetic field for (a) $E_r=1$ kV/m and (b) $E_r=0$ kV/m. The Fermi energy is kept at $E_F=0.035$ eV.}
\label{density}
\end{figure*}
The DOS exhibits oscillation with the magnetic field-known as the SdH oscillation.
The presence of an in-plane electric field is causing a valley separation in SdH oscillation too.
Here, we have taken very weak brodening as $\Gamma_0=0.05\hbar\omega_c$.
Here, we have computed the summation over $k_y$ by using the fact 
that the centre of the cyclotron orbit is always confined within the system
\ie $0\le |x_0+G_{n}|\le L_x$ or $0\le k_{y}\le L_{x}/l_{c}^2$ with 
$G_{n}=\lambda\sqrt{2n}l_c\beta_{\xi}(1-\beta_{\xi})^{-1/4}$ and sebsequently
$\sum_{k_y}\rightarrow\frac{L_y}{2\pi}\int_{0}^{L_x/l_c^2}dk_y=\Omega/2\pi l_c^2$.
Moreover, in the above equation, we have also ignored the $k_y$ dependent
term ($\hbar v_e k_y$) in the LLs expression under the assumption of higher Landau level and weak electric field.
As this term is not associated with valley index, hence this assumption
will not affect the valley dependency of the DOS. The DOS is plotted in Fig.~(\ref{density}) by using Eq.~(\ref{dos}).
It is an established fact\cite{PhysRevB.65.245420,PhysRevB.47.1522}
that the impurity induced LLs broadening in 2D Dirac material is 
directly proportional to $\sqrt{B}$. To plot dimensionless DOS, we consider LLs
broadening width $\Gamma_0=0.05\hbar\omega_c$.

\section{Magnetoconductivity}\label{sec3}
In this section, we evaluate the quantum Hall conductivity and the longitudinal conductivity by using the formalism
based on linear response theory  developed in Ref.~[\onlinecite{charbonneau1982linear}] which has been extensively
used in other 2D systems\cite{PhysRevB.46.4667,PhysRevB.86.115432,PhysRevB.90.235423,tahir2013valley,
tahir2012quantum,tahir2016quantum,Firoz}. In presence of perpendicular magnetic field,
the conductivity becomes a tensor with diagonal $(\sigma_{\mu\nu}^{d})$ as well as non-diagonal 
$(\sigma_{\mu,\nu}^{nd})$ terms \ie $\sigma_{\mu\nu}=\sigma_{\mu,\nu}^{d}+\sigma_{\mu,\nu}^{nd}$, 
where $\{\mu,\nu\}=\{x,y\}$.

\subsection{Quantum Hall conductivity}
The quantum Hall conductivity( $\sigma_{xy}$) of borophene can be evaluated by using the standard formula within
the linear response regime\cite{charbonneau1982linear,PhysRevB.46.4667}:
\begin{eqnarray}
 \sigma_{xy}&=&\frac{ie^2\hbar}{\Omega}\sum_{\zeta\ne\zeta'}f_{\zeta}(1-f_{\zeta'})\la\zeta\mid
 \mathcal{\hat{V}}_x\mid \zeta'\ra
 \la\zeta'\mid\mathcal{\hat{V}}_y\mid\zeta\ra\nonumber\\&\times&\frac{1-\exp{[\beta_T(E_{\zeta}-E_{\zeta'})}]}
 {E_{\zeta}-E_{\zeta'}}\lim_{\eps\to0}\frac{1}{E_{\zeta}-E_{\zeta'}+i\eps}.
\end{eqnarray}
In the above expression, the velocity operators are defined as:
$\mathcal{\hat{V}}_x=\partial \mathcal{H}_0/\partial p_x=\xi v_c\hat{\sigma}_{x}$ and
$\mathcal{\hat{V}}_y=\partial \mathcal{H}_0/\partial p_y=v_{e}^{\xi}\mathbb{1}+\xi v_{c}\hat{\sigma}_{y}$, and
$|\zeta\ra\equiv|\Psi_{\zeta}(x,y)\ra$. Note that here
we have used the transformed momentum operators \ie $p_{x}\rightarrow(v_x/v_y)^{1/2}p_{x}$
and $p_{y}\rightarrow(v_y/v_x)^{1/2}p_{y}$ for the non-tilted part of the Hamiltonian. Also, 
$f_{\zeta}=[1+\exp\{\beta_T(E_{\zeta}-E_{F})\}]^{-1}$ is 
the Fermi-Dirac distribution function with $E_F$ is the Fermi energy and
$\beta_T=(k_BT)^{-1}$ where $k_B$ is the Boltzmann constant.

By using the identity $f_{\zeta}(1-f_{\zeta'})
\{1-\exp{[\beta_T(E_{\zeta}-E_{\zeta'})}]\}=f_{\zeta'}(1-f_{\zeta})$, one can arrive at Kubo-Greenwood
formula for the Hall conductivity in each valley as
 \begin{eqnarray}\label{hc}
 \sigma_{xy}&=&\frac{ie^2\hbar}{\Omega}\sum_{\zeta\ne\zeta'}(f_{\zeta}-f_{\zeta'})
 \frac{\la\zeta\mid \mathcal{\hat{V}}_x\mid \zeta'\ra\la\zeta'\mid\mathcal{\hat{V}}_y\mid\zeta\ra}
 {(E_{\zeta}-E_{\zeta'})^2}.\nonumber\\
\end{eqnarray}
The velocity matrix elements in a particular valley (see Apendix A) are evaluated as (for $n>0$)
\begin{equation}
\la n,k_y\mid \mathcal{\hat{V}}_x\mid n',k_y'\ra=-i\frac{v_c}{2\gamma_{\xi}}
\big[\lambda\delta_{n,n'-1}+\lambda'\delta_{n-1,n'}\big]\delta_{k_y,{k'}_y}
\end{equation}
and 
\begin{eqnarray}
\la n',k_y'\mid \mathcal{\hat{V}}_y\mid n,k_y\ra&=&-(1-\beta_{\xi}^2)\frac{v_c}{2}\nonumber\\&\times&
\left[\lambda'\delta_{n',n-1}+\lambda\delta_{n'-1,n}\right]\delta_{k_y,{k'}_y}.
\end{eqnarray}
The presence of $\delta_{k_y,k_y'}$ guarantees that velocity matrix elements are
non-zero only for $k_y=k_y'$. To proceed further, we now follow the assumption\cite{PhysRevB.83.075427}
that the effects of $k_y$ through Fermi distribution function is very small and hence we can ignore it.
This assumption is well justified for weak electric field. Moreover, it can also be seen that the
$k_y$ dependent term inside the Fermi distribution function is independent of the Landau level index (n),
for which the differences between the two Fermi distribution functions corresponding to the two successive
Landau levels are almost independent of $k_y$.  By using $\sum_{k_y}\rightarrow\frac{L_y}{2\pi}
\int_{0}^{L_x/l_c^2}dk_y=\Omega/2\pi l_c^2$ in Eq.~(\ref{hc}), we obtain (for $n>0$)
\begin{eqnarray}\label{kubo}
 \sigma_{xy}^{\xi}&=&\frac{e^2}{h}\sum_{n}\frac{(f_{n,\xi}-f_{n+1,\xi})}
 {(\sqrt{2n}-\sqrt{2(n+1)})^2}.
\end{eqnarray}
\begin{figure}[htb]
{ \includegraphics[width=.5\textwidth,height=6cm]{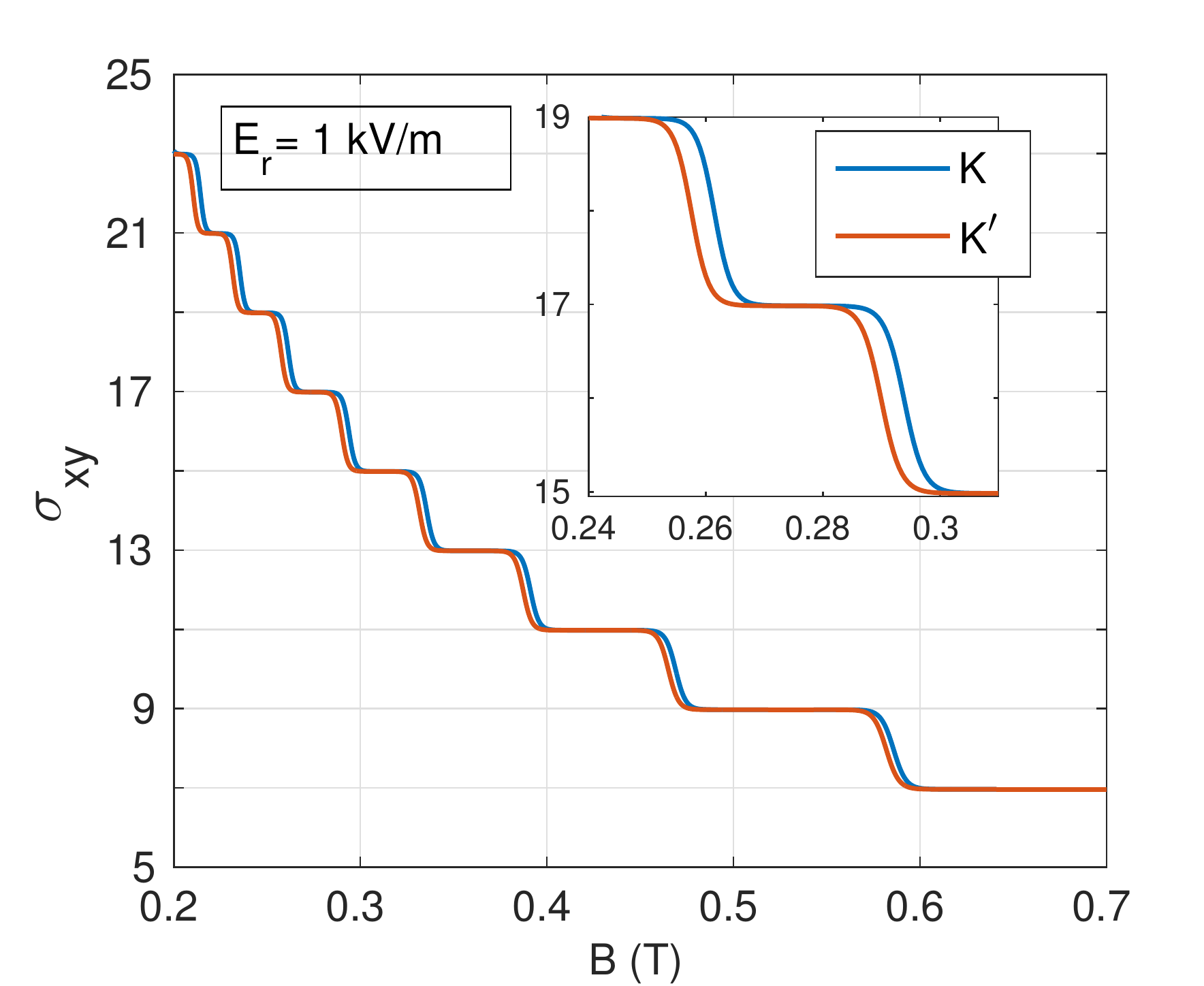}}
\caption{Quantum Hall conductivity versus magnetic field. The Hall conductivity is in units of $e^2/h$.
The tilt velocity $v_t=0.32$ unit, the Fermi energy $E_{F}=0.037$ eV and the temperature is taken much low at $T=0.1$K.}
\label{qhc}
\end{figure}
At zero temperature, if $E_{F}$ lies between $\vareps_{n}$
and $\vareps_{n+1}$-th Landau level, then above expression can be reduced to
\begin{equation}\label{hall2}
 \sigma_{xy}^{\xi}=2\frac{e^2}{h}\left(n+\frac{1}{2}\right),
\end{equation}
including the spin degeneracy but without valley degeneracy. The contribution which arises from $n=0$
level has to be evaluated separately. The velocity matrix elements between $n=0$ and $n'$ are given by
\begin{equation}
\la 0,k_y\mid \mathcal{\hat{V}}_x\mid n',k_y'\ra=-i\frac{v_c}{2\gamma_{\xi}}\lambda\delta_{0,n'-1}\delta_{k_y,{k'}_y}
\end{equation}
and
\begin{eqnarray}
\la n',k_y'\mid \mathcal{\hat{V}}_y\mid 0,k_y\ra&=&-(1-\beta_{\xi}^2)\frac{v_c}{2}
\lambda\delta_{n'-1,0}\delta_{k_y,{k'}_y}.
\end{eqnarray}
Finally, the Hall conductivity due to zero-th Landau level is
\begin{eqnarray}
 \sigma_{xy}^{0}&=&\frac{e^2}{h}\sum_{\xi}(f_{0,\xi}-f_{1,\xi}).
\end{eqnarray}
The quantization of Hall conductivity, in Eq.~(\ref{hall2}), is
exactly similar to the case of graphene without valley degeneracy i.e,., $\sigma_{xy}=1,3,5,7,9...$ in units of $e^2/h$.
However, the valley dependency appears at the Hall steps governed by the valley dependent Fermi-distribution function.
The Hall conductivity obtained in Eq.~(\ref{kubo}) is plotted numerically in Fig.~(\ref{qhc}).
The Fermi energy is taken to be $E_{F}=0.037$ eV, corresponds to carrier density
$n_e=10^{15}m^{-2}$ and tilt velocity $v_t=0.32$ unit. The Fermi energy can be evaluated 
numerically in terms of magnetic field for a particular carrier density, as
done in Ref.~(\onlinecite{Firoz_weiss}).\\
The quantum Hall conductivity plots in Fig.~(\ref{qhc}) show a series of unequal quantum Hall 
plateaus, as expected. However, most importantly, two valleys are not following the same steps
although exhibiting identical plateaus. The valley separation around the steps are exclusively
caused and governed by the in-plane electric field i.e., Hall field. This feature is in complete
contrast to the case of monolayer graphene under the influences of the Hall field. The origin of the valley
separation at the steps can be traced to the lifting of the valley degeneracy in presence of an 
in-plane electric field in the Landau levels of borophene, where as in graphene such removal of valley
degeneracy does not occur.
\subsection{Longitudinal conductivity}
In this subsection, we investigate the longitudinal conductivity. In general, the longitudinal conductivity arises
mainly due to the scattering of cyclotron orbits from the charge impurities.
This contribution is also known as collisional conductivity. In low temperature regime, scattering mechanism
can be treated as elastic on the ground that charge carriers can not offer enough energy 
to excite charge impurity from its ground states to excited states during collisions. First we
consider the case of the presence of a Hall field.
\subsubsection{In presence of Hall field}
The collisional conductivity in low temperature regime can be evaluated by using the 
following formula\cite{charbonneau1982linear,PhysRevB.46.4667}
\begin{equation}\label{coll}
\sigma_{xx}=\frac{\beta_T e^2}{2\Omega }\sum_{\zeta,\zeta'}f_{\zeta}(1-f_{\zeta'})
W_{\zeta,\zeta'}(x_{\zeta}-x_{\zeta'})^2.
\end{equation}
Here, $x_{\zeta}=\la\zeta\mid x \mid \zeta\ra$ is the average value of the x-component of the
position operator of an electron in state  $\mid \zeta\ra$, which can be evaluated to be
$x_0+G_{n}$ (=$k_yl_c^2+G_{n}$). To proceed further analytically, we can drop the $n$ dependent term 
($G_n$) in the centre of the cyclotron orbit and thus $(x_{\zeta}-x_{\zeta'})^2=(q_yl_c^2)^2$
with $k_y'-k_y=q_y$. Note that dropping of $G_n$ would not make any drastic changes in the main
result except a small effect to the conductivity amplitude. Moreover, for intra Landau level scattering, $G_n$ would
get canceled out automatically in the expression of $(x_{\zeta}-x_{\zeta'})$. The key features of the longitudinal
conductivity oscillations is preserved in the $n$-dependent part of the Landau levels in the Fermi distribution function,
which controls the oscillations.
 \begin{figure}[htb]
 {\includegraphics[width=.5\textwidth,height=6cm]{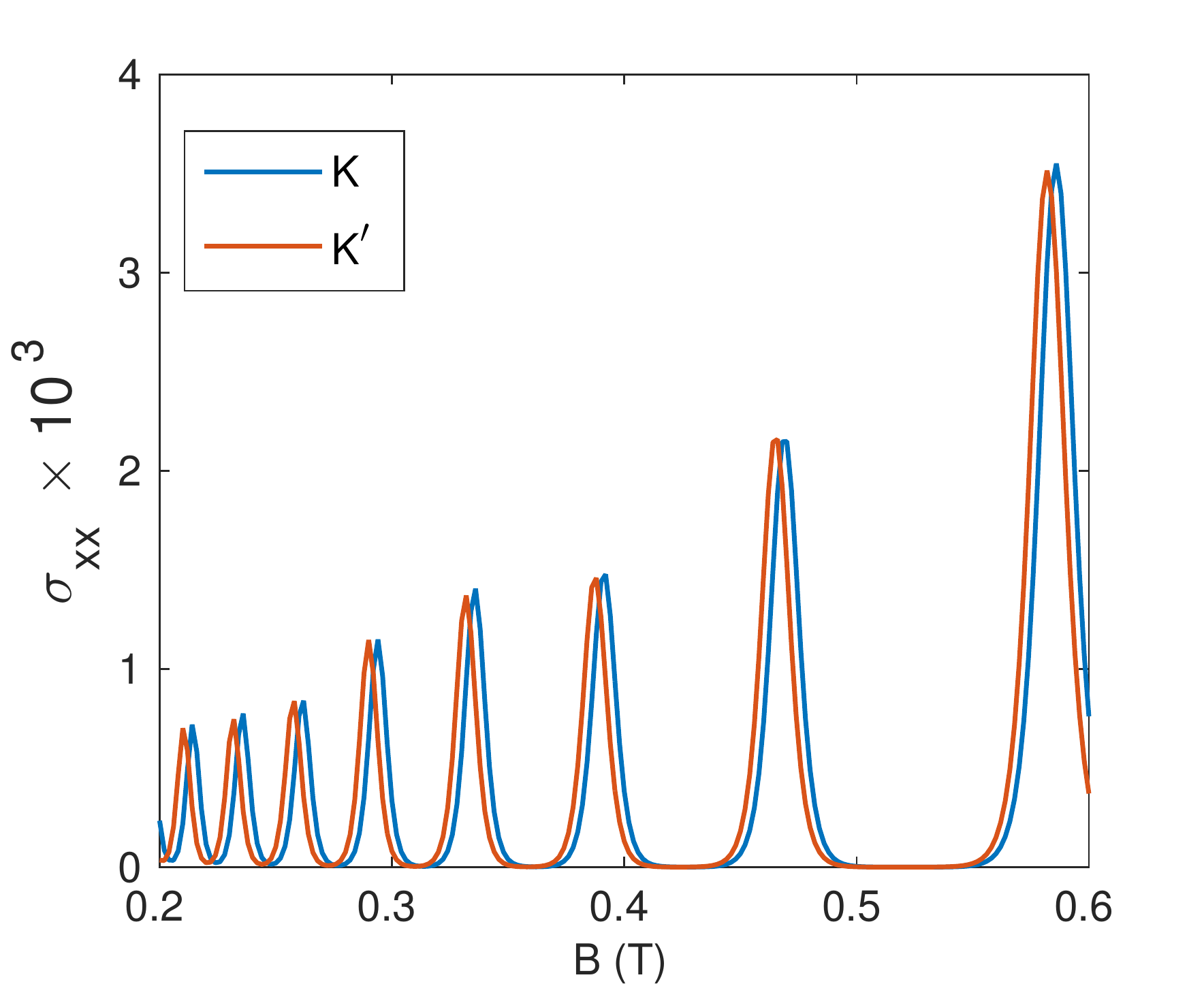}}
 \caption{Longitudinal conductivity (in units of $e^2/h$) versus magnetic field. Other parameters
 are as: the Fermi energy $E_{F}=0.037$ eV, the Temperature $T=3$ K, the impurity density
 $n_i=10^{13}m^{-2}$ and LLs broadening $\Gamma_0=0.1\hbar\omega_c$.}
 \label{long_con}
 \end{figure}
On the other hand, the scattering rate between states $|\zeta\ra$ and $|\zeta'\ra$ is given by
\begin{equation}
 W_{\zeta,\zeta'}=\frac{2\pi n_i}{\Omega\hbar}\sum_{q}\mid U_q\mid^2\mid F_{\zeta,\zeta'}
 (\eta)\mid^2\delta(E_{\zeta}-E_{\zeta'})\delta_{k_y,k_y'+q_y}.
 \end{equation}
Here, $n_i$ is the impurity density and $\eta=q^2l_c^2/2$. The $2$D Fourier transformation of the
screened charged impurity potential $U(r)=[e^2/{4\pi \epsilon_0\epsilon_r r}]e^{-k_sr}$ is
$ U_q=U_0[q^2+k_s^2]^{-1/2}\simeq U_0/k_s$ for short range delta function-like potential, where
$U_0=e^2/(2\epsilon_0\epsilon_r)$ and $k_s$ is the screening vector. The form factor is defined
as $F_{\zeta,\zeta'}(\eta)=\la\zeta\mid e^{i\vec{q}.\vec{r}}\mid \zeta'\ra$, which can be evaluated
(See Apendix B) considering
only $n'=n\pm1$ (because of the presence of $\delta_{k_y,k_{y'}+q_y}$ in $W_{\zeta,\zeta'}$)
\begin{equation}
 \mid F_{n,n\pm1}(\eta)\mid^2=e^{-\eta}[R_{n,n\pm1}(\eta)]^2
\end{equation}
with 
\begin{equation}
R_{n,n+1}(\eta)\simeq\frac{1}{\sqrt{2(n+1)}}L_{n}^{1}(\eta)+\frac{1}{\sqrt{2n}}L_{n-1}^{1}(\eta)
\end{equation}
and 
\begin{equation}
R_{n,n-1}(\eta)\simeq\sqrt{\frac{2}{n}}L_{n-1}^{1}(\eta)+\sqrt{\frac{2}{n-1}}L_{n-2}^{1}(\eta).
\end{equation}
Here, $L_{n}(\eta)$ is the Laguerre polynomial of order $n$. 
For lowest LL ($n=0$), the scattering amplitude has to be evaluated separately as
\begin{equation}
  \mid F_{0,1}(\eta)\mid^2=e^{-\eta}\frac{1}{2}[L_{0}^{1}(\eta)]^2.
\end{equation}
By replacing summation over $k_y$ by $\Omega/(2\pi l_c^2)$,
 $ \sum_{q}\rightarrow\frac{\Omega}{(2\pi)^2}\int q dq d\phi$ and $(x_{\xi}-x_{\xi'})^2=q_y^2l_c^4=(q\sin\phi)^2l_c^4$ ,
the Eq.~(\ref{coll}) can be further simplified to $\sigma_{xx}=\sum_{\xi}\sigma^{\xi}$ with
\begin{equation}\label{cond}
\sigma_{xx}^{\xi}\simeq \frac{e^2}{h}\frac{n_{i}U_s^2}{2\pi l_c^2\Gamma_0}\beta_{T}\sum_{n,\xi}I_nf_{n,\xi}(1-f_{n,\xi}).
\end{equation}
Here, $U_s=U_0/k_s$, $\Gamma_0$ is the impurity induced LLs broadening and 
\begin{equation}
I_{n}=\int_{0}^{\infty}\eta^2 e^{-\eta}\left([R_{n,n+1}(\eta)]^2+[R_{n,n-1}(\eta)]^2\right) d\eta.
\end{equation}
The cut-off limit of the above integration can be extracted from the short range scattering condition \ie $q<<k_s$ 
($\eta<<\eta_s$ with $\eta_s=k_s^2l_c^2/2$). 

We plot longitudinal conductivity in Fig.~(\ref{long_con}) by using Eq.~(\ref{cond}). 
For this numerical plot, we use the following parameters: charge density $n_e=10^{15}m^{-2}$,
impurity density $n_i=10^{13} m^{-2}$, temperature $T=3$ K, dielectric constant of borophene
is taken to be $\kappa=10$ which is in consistent with Ref.~[\onlinecite{PhysRevB.96.035410}]
and screening vector $k_s=10^{8}m^{-1}$. We also consider the magnetic field dependency of the 
impurity induced LLs broadening as $\Gamma_0\simeq 0.1\hbar\omega_c$. As it is proven
fact\cite{PhysRevB.46.4667,PhysRevB.81.085402} that SdH oscillation start to die out
with the increasing temperature, hence we give plots only for a particular temperature.
\begin{figure*}
\subfigure[]
{\includegraphics[width=.49\textwidth,height=6cm]{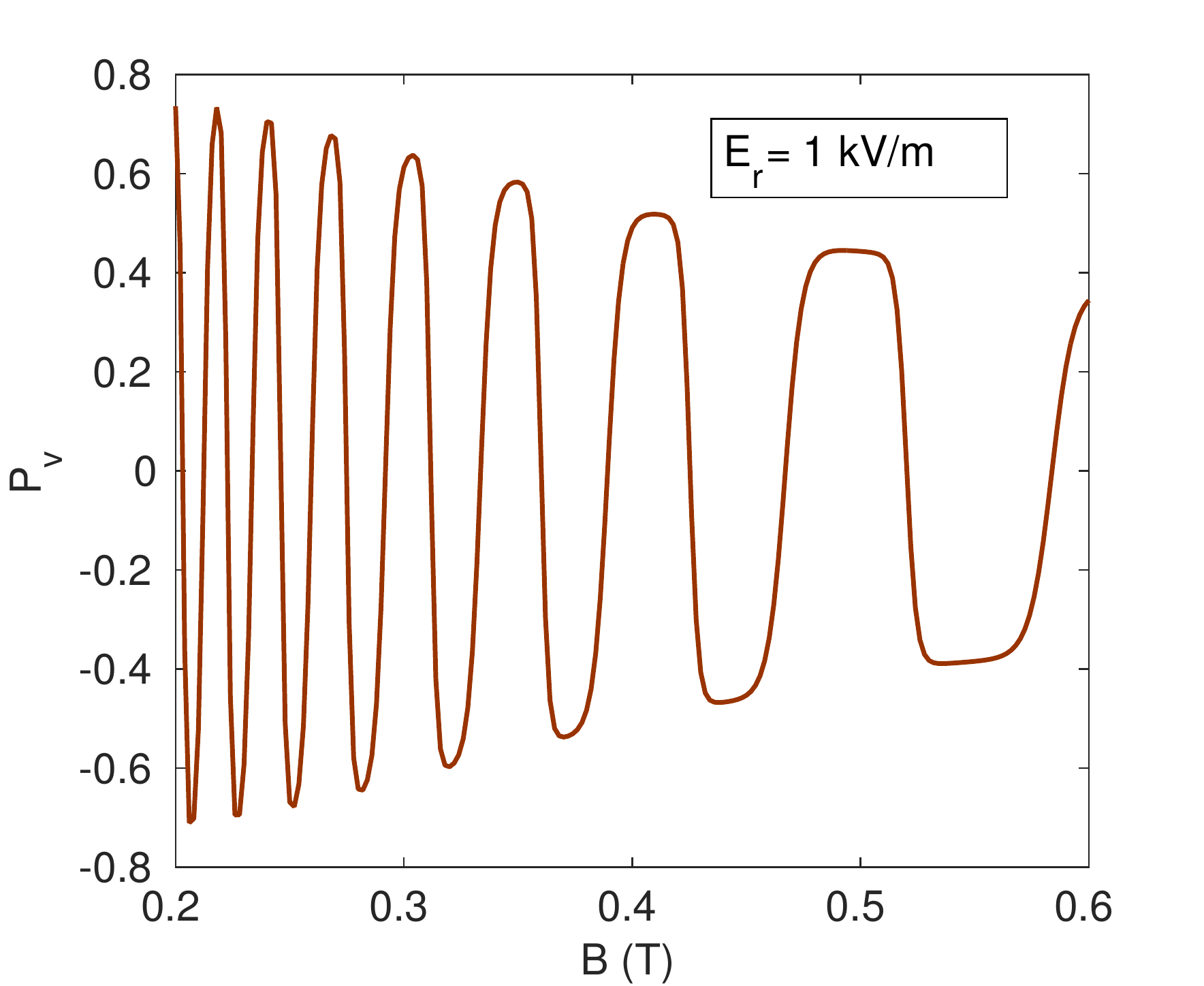}}
\subfigure[]
{\includegraphics[width=.49\textwidth,height=6cm]{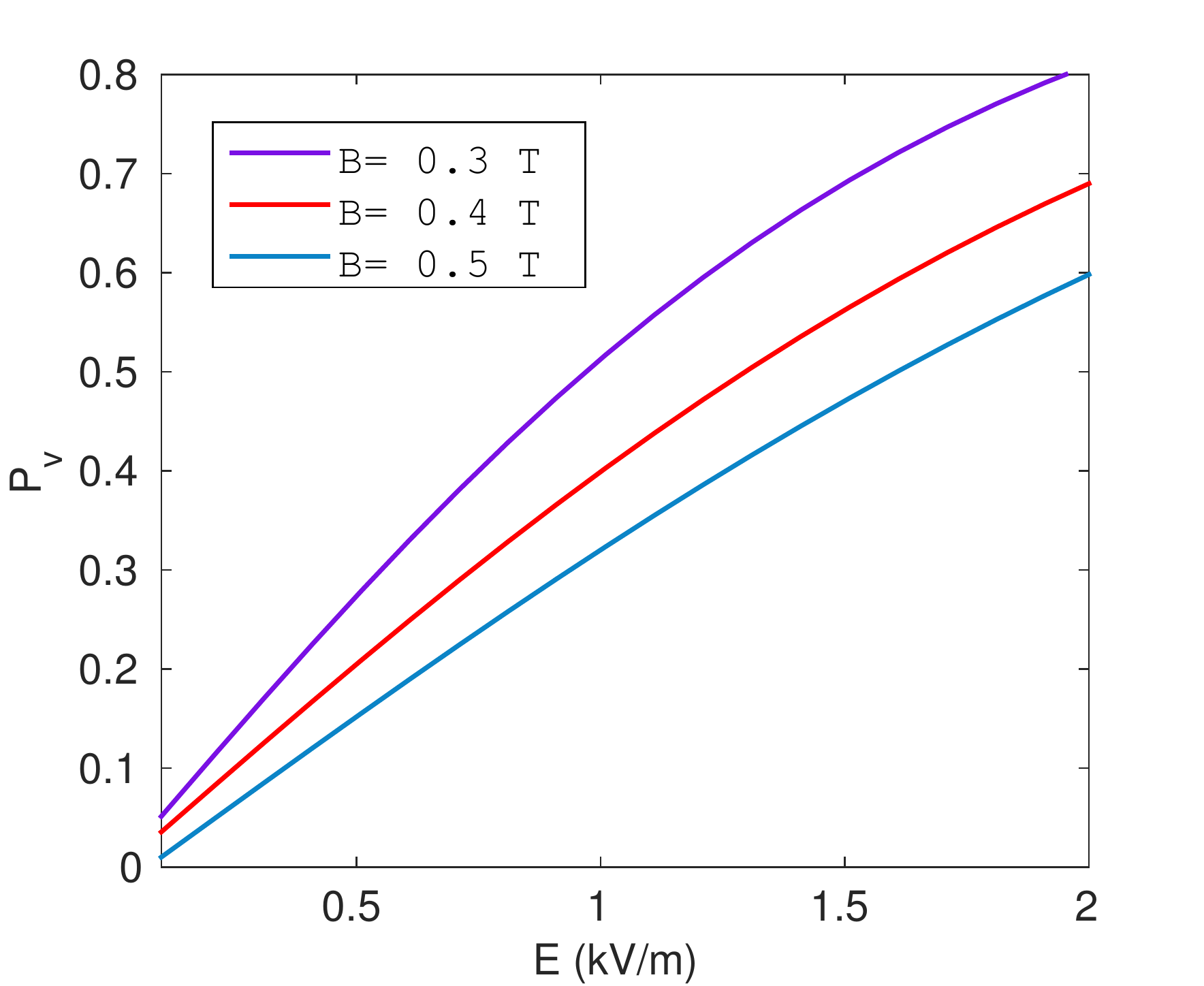}}
\caption{Polarization in the longitudinal conductivity versus (a) magnetic field and (b) electric field.}
\label{pola_elec}
\end{figure*}
The longitudinal conductivity peaks are corresponding to the crossing of Fermi level through the LLs.
However, because of the valley separation in LLs, conductivity peaks in two valleys are not
at the same location. The separation of the conductivity peaks in two valleys are the direct
consequences of the lifting of the valley degeneracy in the LLs. The gap between two consecutive
peaks in each valley increases with the increase of magnetic field and this is obvious as
the LLs spacing  between two successive LLs also increases with the magnetic field.\\

In Fig.~(\ref{pola_elec})a, we plot the polarization in the longitudinal conductivity
versus magnetic field and electric field by using the relation
\begin{equation}
P_{\rm v}=\frac{\sigma_{xx}^{+}-\sigma_{xx}^{-}}{\sigma_{xx}^{+}+\sigma_{xx}^{-}}.
\end{equation}
It shows that a sizable valley polarization
can be achieved for a wide range of magnetic field by applying a Hall field.
However, the polarization appears to be oscillatory with magnetic field.
This is because of the oscillatory nature of
longitudinal conductivity with magnetic field in both valleys. On the other hand, we also
show the evolution of polarization with electric field for three values of magnetic field in
Fig.~(\ref{pola_elec})b which shows that a sizable polarization can emerge for 
an electric field $E>0.5$ kV/m. Note that although a weak fluctuation of Fermi energy
between nearest LLs does exist with respect to the magnetic field [see Ref.~(\onlinecite{Firoz_weiss})],
we keep Fermi level constant in the regime of interest as the amplitude of
Fermi energy fluctuation is very small.
\begin{figure}[!thpb]
\centering
\includegraphics[height=6cm,width=0.90 \linewidth]{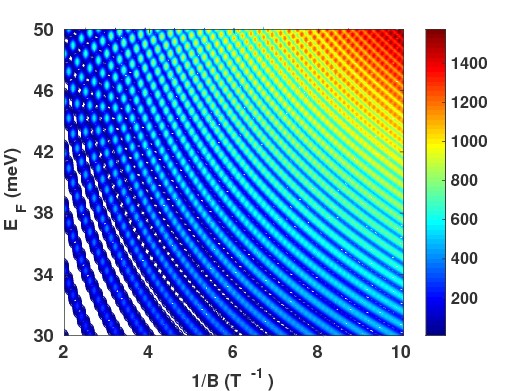}
\caption{(Color online) Contour plot of longitudinal conductivity (in units of $e^2/h$)
in the plane of $1/B$ and $E_{F}$. The temperature is taken as $T=3$ K.}
\label{sdh_contour}
\end{figure}
\subsubsection{In absence of Hall field}
In this subsection, we evaluate the longitudinal conductivity in absence of the Hall field.
In absence of the Hall field ($E_{r}=0$), the LLs regain the valley degeneracy and hence the valley dependent 
transport is not expected anymore. The scenario is now quite similar to the case of the monolayer graphene
without Hall field, except the renormalized Fermi velocity $v_c=\sqrt{v_xv_y}$ and tilting
of the Dirac cones. Moreover, in absence of the Hall field the LLs not only recover the 
valley degeneracy but also get back the $k_y$ degeneracy for which the intra-LLs scattering
is now allowed, and the scattering rate between states $|\zeta\ra$ and $|\zeta'\ra$ is now given by
\begin{equation}
 W_{\zeta,\zeta'}=\frac{2\pi n_i}{\Omega\hbar}\sum_{q}\mid U_q\mid^2\mid Q_{\zeta,\zeta'}
 (\eta)\mid^2\delta(E_{\zeta}-E_{\zeta'})\delta_{k_y,k_y'+q_y}.
 \end{equation}
 Here,
 \begin{equation}
 \mid Q_{n,n}(\eta)\mid^2=\frac{e^{-\eta}}{4}[J_{n,n}(\eta)+J_{n-1,n-1}(\eta)]^2
\end{equation}
Following the Ref.~[\onlinecite{PhysRevB.86.115432,PhysRevB.81.085402}],
we obtain the longitudinal conductivity as
\begin{equation}\label{cond2}
\sigma_{xx}\simeq \frac{e^2}{h}\frac{n_{i}U_s^2}{\pi l_c^2\Gamma_0}\beta_{T}\sum_{n}nf_n(1-f_n)
\end{equation}
which is plotted in Fig.~\ref{sdh_contour} in the plane of Fermi level and inverse magnetic field.
The longitudinal conductivity shows SdH oscillations with inverse magnetic field
as well as Fermi level both with different frequencies. The conductivity amplitude increases with 
the Fermi level as well as inverse magnetic field. However, to understand the effect of tilt parameter
on SdH oscillations frequency, we plot the longitudinal conductivity versus inverse magnetic field for both 
cases \ie in absence and presence of tilt parameter in Fig.~\ref{sdh}.
\begin{figure}[!thpb]
\centering
\includegraphics[width=.49\textwidth,height=6cm]{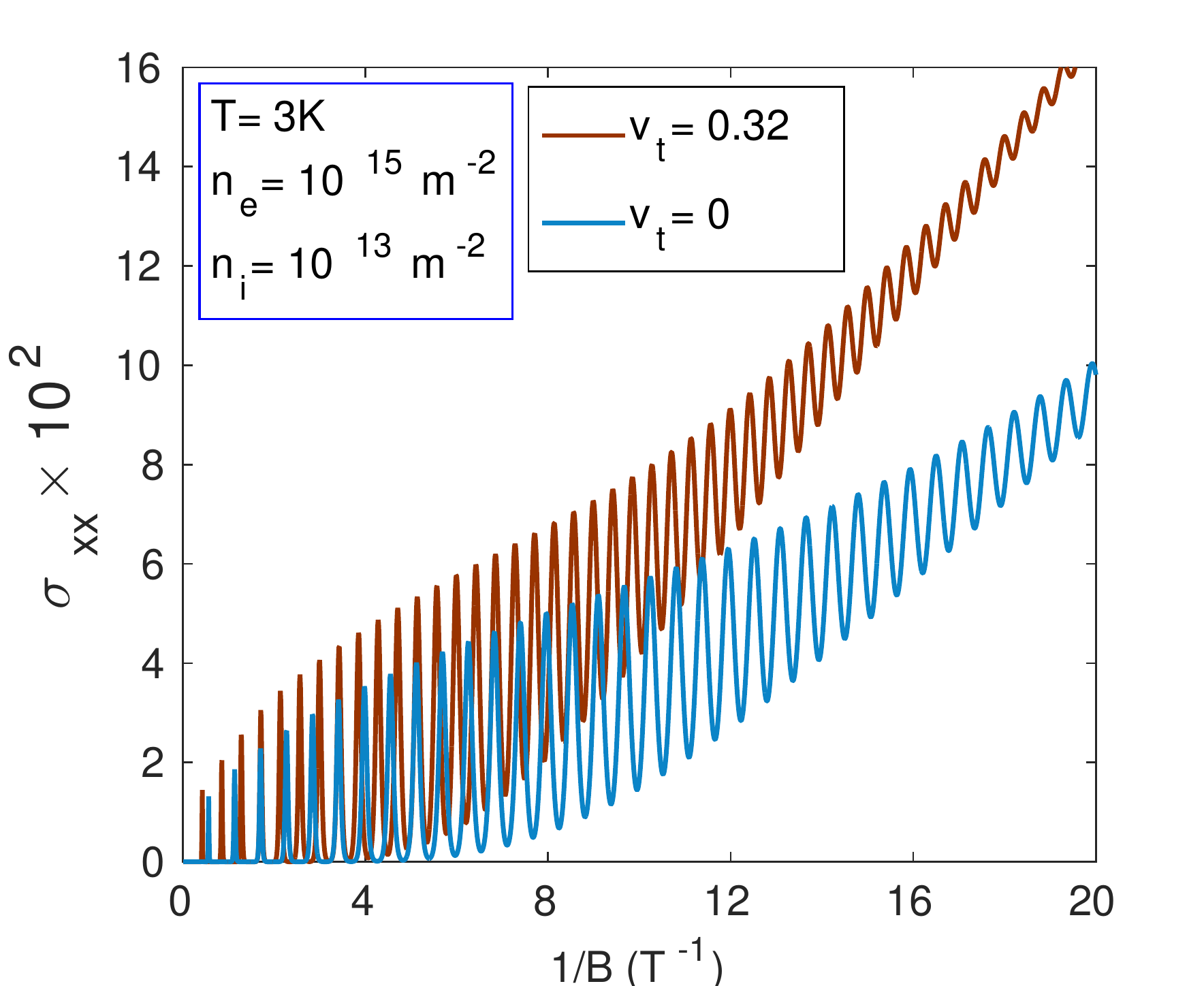}
\caption{(Color online) Longitudinal conductivity (in units of $e^2/h$) versus inverse magnetic field.
The Fermi level $E_{F}= 0.037$ eV.}
\label{sdh}
\end{figure}
It is observed that the frequency as well as the amplitude of the longitudinal conductivity are significantly 
affected by the tilt parameter. To get a clear picture of the influences of the tilt parameter on
the SdH oscillations, an approximate analytical simplification of the longitudinal conductivity
is necessary. To obtain analytical expression of Eq.~(\ref{cond2}),
we replace the summation over LL index $n$ as: $\sum_{n}\rightarrow \pi l_c^2\int_0^{\infty}D(E)dE$,
where $D(E)$ is the density of states (DOS). The analytical approximate form of the DOS can be 
obtained by following the Refs.[\onlinecite{poisson_peeters,PhysRevB.90.205417,Firoz_JPCM}] as
\begin{equation}
 D(E)\simeq\frac{2E}{\pi (\hbar v_c)^2}\left\{1+2\Omega(E)
 \cos\left[2\pi\left(\frac{E}{\hbar\bar{\omega}_c}\right)^2\right]\right\},
\end{equation}
where impurity induced damping factor is $\Omega(E)=
\exp\left\{-2\pi\left[\frac{\Gamma(E)E}{(\hbar\bar{\omega}_c)^2}\right]\right\}$
with $\Gamma(E)=4\pi\Gamma_0^2 E/(\hbar\bar{\omega}_c)^2$ and $\bar{\omega}_c=\omega_c/(1-\beta^2)^{3/4}$
with $\beta=v_t/v_c$. 
Using the above form of DOS in Eq.~(\ref{cond2}), one can readily find
\begin{equation}\label{collision}
 \sigma_{xx}\simeq\frac{e^2}{h}U\left\{1+2\Omega(E_F)\frac{T/T_c}{\sinh(T/T_c)}
 \cos\left[2\pi\left(\frac{f}{B}\right)\right]\right\}.
\end{equation}
Here, $U$ is a dimensionless factor and given by 
\begin{equation}\label{ampli}
 U=\frac{n_iU_s^2}{2\Gamma_0}\left[\frac{E_F}{\hbar\bar{\omega}_c}\right]^2\frac{E_F}{\pi(\hbar v_c)^2}.
\end{equation}
On the other hand, the SdH oscillations frequency with the inverse magnetic field is given by
\begin{equation}\label{fre}
f=\frac{1}{e\hbar}\left(\frac{E_F}{v_c}\right)^2g_t,
\end{equation}
where $g_t=1/(1-\beta^2)^{3/2}>1$. The above expression shows that the tilt parameter amplifies
the frequency of the SdH oscillation by a factor $g_t=1.1526$. The characteristic temperature
is defined by $k_BT_c=(\hbar\bar{\omega}_c)^2/(4\pi^2 E_F)$, beyond which the SdH oscillation start to die out. 
Note that apart from the frequency, the characteristic temperature is also affected by the tilt
parameter.

In addition to the inverse magnetic field, longitudinal conductivity also exhibits similar SdH
oscillation with the Fermi energy. The SdH oscillations frequency with Fermi energy is 
\begin{equation}
\bar{f}=\frac{E_{F}}{(\hbar\omega_c)^2}g_t.
\end{equation}
This expression shows that unlike the frequency of SdH oscillations with inverse magnetic field [see Eq.~(\ref{fre})],
SdH oscillations with the Fermi energy is non-periodic as the frequency itself depends on the Fermi level strongly.
The tilted parameter suppresses SdH oscillation frequency in both cases in similar fashion. 
The longitudinal conductivity shows the SdH oscillations with the inverse magnetic field and the Fermi energy both.

\section{Effects of strain}
In this section, we investigate how the application of strain can influence the magnetoconductiity
in absence and presence of an in-plane electric field. The issue of strain induced valley 
polarization in Dirac material is not new rather it has been extensively considered in
graphene\cite{PRL_neto,PhysRevB.80.045401,PhysRevB.87.121408} to silicene\cite{PhysRevB.97.085427,
PhysRevB.96.205416}. In Ref.~[\onlinecite{PhysRevB.94.165403}], strain induced quantum valley Hall
effect was also predicted.
However, in this work, we would like to consider the case of interplay 
between the in-plane electric field and the strain. The strain can be described by the displacement field
$\mathcal{U}(r)$ and the strain tensor $\mathcal{U}_{ij}=\frac{1}{2}[(\frac{\partial \mathcal{U}_i}{\partial r_j})
+(\frac{\partial \mathcal{U}_j}{\partial r_i})]$. The strained borophene must not violate any symmetry
possessed by the material. Similar to the graphene, the strain acts as a pseudo magnetic field.
However, the strain induced vector potential in two valleys are in opposite sign and it can be captured
in low energy effective Hamiltonian in presence of a real magnetic field (B) as
\begin{equation}
 H=\xi[v_xp_x\sigma_x+v_y(p_y+eA_{S})\sigma_y+v_t(p_y+eA_{S})\mathbb{1}].
\end{equation}
Here, the vector potential $A_{S}=x(B+\xi t)$ with the strain induced 
pseudo magnetic field $t=[(\frac{\partial {A}_y}{\partial x})-(\frac{\partial {A}_x}{\partial y})]$. Here, the 
different components of vector potential can be expressed\cite{PhysRevB.94.165403} as 
$ {\bf A}=[\alpha_{xy}\mathcal{U}_{xy},\alpha_{xx}\mathcal{U}_{xx}+\alpha_{yy}\mathcal{U}_{yy}$] with
$\alpha_{xy}=3.86$ G-cm, $\alpha_{xx}=3.58$ G-cm and $\alpha_{yy}=-1.15$ G-cm. The strain field in
8-Pmmn borophene is taken to be as $\mathcal{U}(r)=(0,x^2/L,0)$. The strain induced magnetic field
can be estimated to be around $t=100$ T for the sample length of $10$ nm. This is in fact an advantage
of strain that one can generate very large magnetic field. However, in our case we shall keep the strain induced
magnetic field much smaller in order to compete it with the real magnetic field.
The Landau levels of strained borophene in absence of the in-plane electric field can be 
obtained following the Sec.(\ref{sec2}). as
\begin{equation}
 E_{\zeta}=\lambda\hbar\omega_\xi\sqrt{2n}(1-\beta^2)^{3/4},
\end{equation}
where $\omega_{\xi}=\sqrt{v_xv_y}/l_{\xi}$ with $l_{\xi}=\sqrt{\hbar/e(B+\xi t)}$. The Landau levels
of strained borophene is sensitive to the valley index, which is in fact similar to the case of graphene.
On the other hand if we switch-on the in-plane electric field the Landau levels becomes
\begin{equation}\label{ll_str}
 E_{\zeta}=\lambda\hbar\omega_\xi\sqrt{2n}(1-\beta_\xi^2)^{3/4}-\hbar k_y \left[\frac{E_r}{B+\xi t}\right],
\end{equation}
where the valley dependency is now attributed to two different origin. One is the valley
dependent cyclotron frequency arises from the effect of strain and other one is the in plane 
electric field induced valley dependency of $\beta_{\xi}=[E_r/(B+\xi t)+\xi v_t]/\sqrt{v_xv_y}$.
The critical electric field, needed for  LLs in each valley to get collapsed, in presence of strain
would be also modified as $E_{r}=(B+\xi t)(v_c+\xi v_t)$. The quantum Hall conductivity 
corresponding to each valley are plotted in Fig.~(\ref{all_hall}) for 
different combination of strain and electric field. In Fig.~(\ref{all_hall})a, we show that 
both valley are showing identical plateaus as expected in absence of any in-plane electric field and strain.
The effect of electric field is shown again here in Fig.~(\ref{all_hall})b, showing small
separation between two valleys in its steps. The inclusion of strain is plotted in Fig.~(\ref{all_hall})c,
which shows that the valley separation is very large and dominated by the effect of the valley dependent
cyclotron frequency in LLs. Finally, we show only the effect of strain without any in-plane electric field
in Fig.~(\ref{all_hall})d which is very similar to Fig.~(\ref{all_hall})c confirming the 
fact that strain alone can dominate the valley polarization even in presence of a Hall field.
\begin{figure}[!thpb]
\centering
\includegraphics[height=6cm,width=0.98 \linewidth]{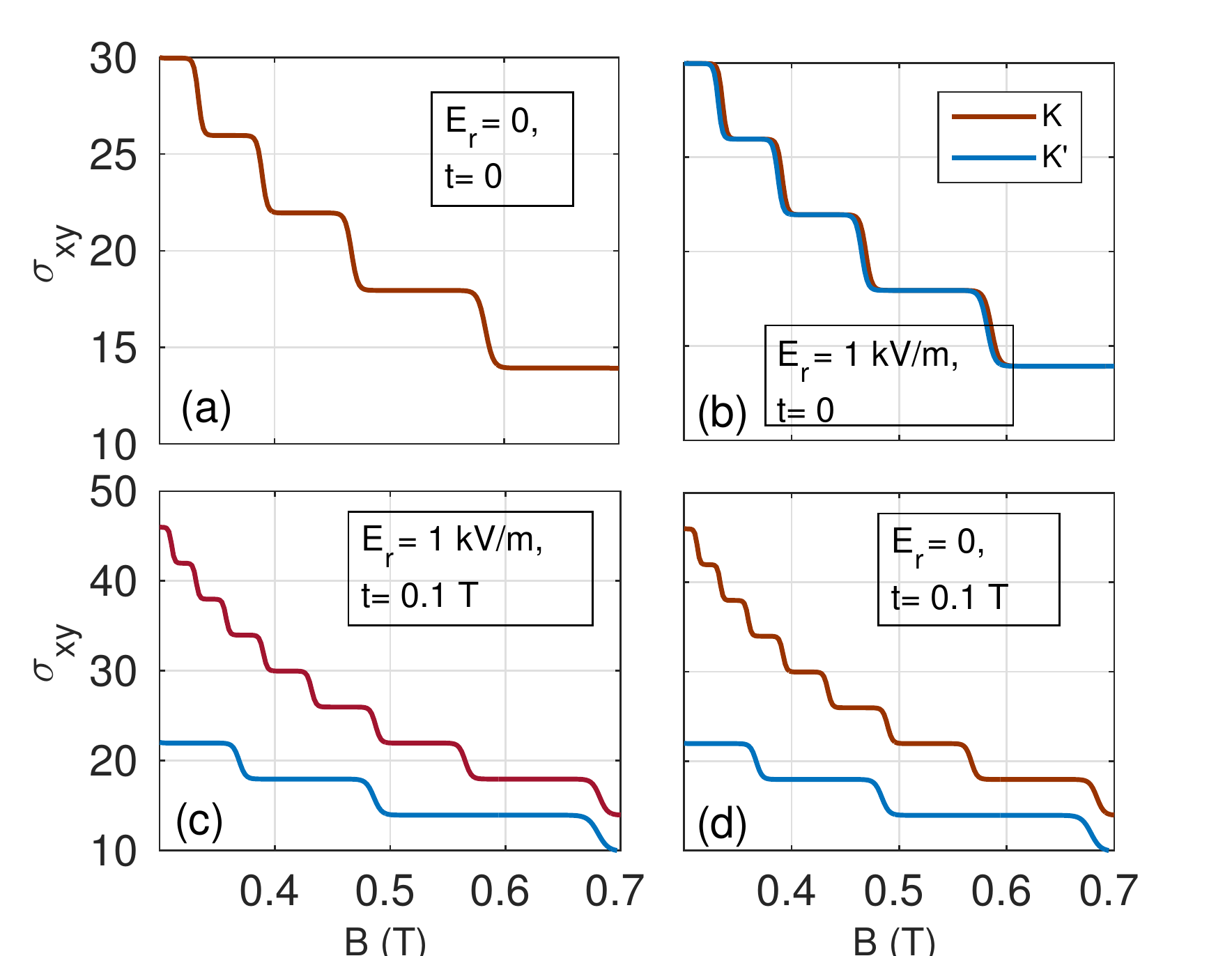}
\caption{(Color online) Plots of quantum Hall conductivity (in units of $e^2/h$)
versus magnetic field. The temperature is taken as $T=2$ K.}
\label{all_hall}
\end{figure}
\begin{figure}[htb]
{ \includegraphics[width=.5\textwidth,height=6cm]{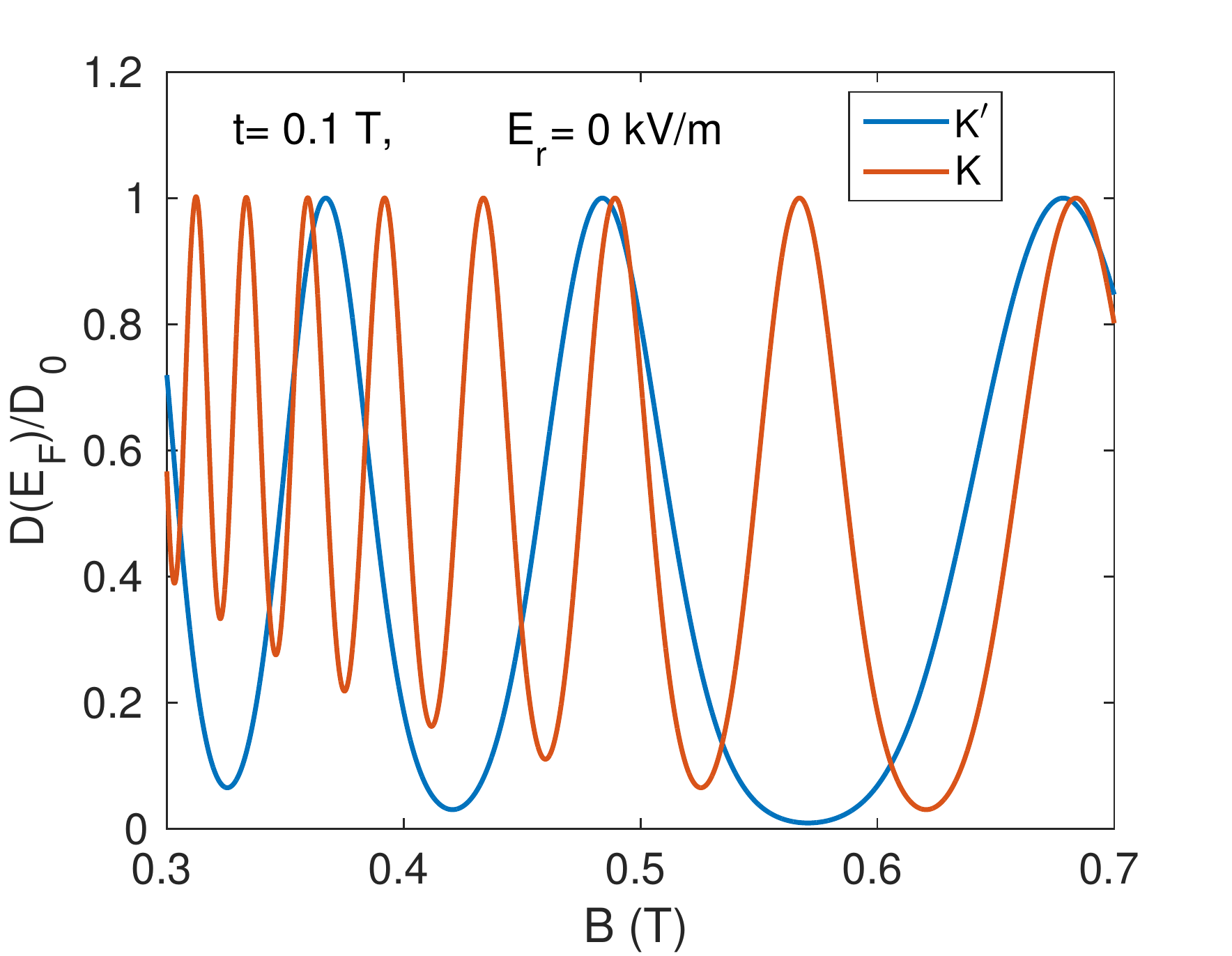}}
\caption{Density of states iin presence of strain versus real magnetic field.
The tilt velocity $v_t=0.32$ unit and the Fermi energy $E_{F}=0.037$ eV.}
\label{dos_dtrain}
\end{figure}
Finally, we shall discuss the case of an in-plane electric field and strain without any 
real magnetic field. From the above discussion, we can write the Landau levels 
in absence of the real magnetic field but in presence of the pseudo magnetic field (strain)
by just setting $B=0$ in Eq.~(\ref{ll_str}). However, the LLs in presence of pseudo magnetic field
and a in-plane electric field will be identical to Eq.~(\ref{ll_str}) except the tilt dependent factor 
$\beta_\xi=\xi\beta$ with $\beta=(E_r/t+v_t)/\sqrt{v_xv_y}$ leading to the LLs as 
\begin{equation}\label{ll_str}
 E_{\zeta}=\lambda\hbar\omega\sqrt{2n}(1-\beta^2)^{3/4}-\xi\hbar k_y \left(\frac{E_r}{t}\right).
\end{equation}
Here, the valley index is not intrinsically associated with the LLs index $n$,
for which valley dependent magnetotransport is not expected under our assumption of weak
electric field. Hence, we can conclude that the in-plane electric field looses its ability
to produce valley dependent Hall steps in quantum Hall conductivity if real magnetic field 
is replaced by strain induced pseudo magnetic field. 

The difference in the size of plateaus in presence of strain, as shown in Fig.~(\ref{all_hall})d, can also be
explained from the SdH oscillation feature in DOS plots in Fig.~(\ref{dos_dtrain}). Here, both valley exhibit 
large frequency differences in SdH oscillation, which reflects as unequal plateaus in Hall conductivity too.

\section{Summary and conclusions}\label{sec4}
In this work, we have studied the magnetotransport properties of a 2D sheet of the polymorph 
of $8$-$Pmmn$ borophene which exhibits tilted anisotropic Dirac cones in its band structure.
We have applied an in-plane electric field (Hall field) to remove the valley degeneracy in its LLs.
The signatures of the lifting of the valley degeneracy in the LLs are examined in the magnetotransport properties.
We have evaluated the quantum Hall and the longitudinal conductivity in presence of the Hall field 
by using linear response theory. The presence of the Hall field causes valley dependent
longitudinal and Hall conductivity, which is in complete contrast to the case of
monolayer graphene\cite{PhysRevB.83.075427} where the Hall field does not 
remove the valley degeneracy from its LLs. A sizable valley polarization can be 
achieved in the longitudinal conductivity by applying a Hall field which can be linked to the 
field of valleytronics where valley dependent transport by tuning external parameter 
is one of the the key requirement. Here, it is worthwhile to mention that 
the external in-plane electric field should be along the direction of tilt induced pseudo
electric field. If external real electric field is taken perpendicular to the direction
of pseudo electric field then exact solution is not possible and numerical study may
not yield valley polarization as the real electric field will affect almost equally to both valleys. 

Moreover, we have also noted by analyzing analytical results that SdH oscillation
frequency is enhanced by the tilting of the Dirac cones.
Finally we have also discussed the possible scenario if the real magnetic field
is replaced by a strain induced pseudo magnetic field and found that in this case Hall field
can not lift the valley degeneracy. However, if the real and pseudo magnetic field both present then
the lifting of the valley degeneracy would be dominated by the strain induced pseudo magnetic field.
\begin{appendix}
\section{}
The velocity matrix elements are given by
\begin{eqnarray}\label{matrix1}
 &&\la n,k_y|\hat{\mathcal{V}}_x|n',k'_y\ra= \xi v_c\la n|\hat{\sigma}_x|n'\ra\delta_{k_y,k_y'}\nonumber\\
 &&=\xi\frac{v_c}{2\gamma_{\xi}} (\lambda' A^{\dagger}\phi_n+i\xi B^{\dagger}\phi_{n-1})
 \sigma_x(\lambda A\phi_{n'}+i\xi B\phi_{n'-1})\nonumber
\end{eqnarray}
where
\begin{equation}
 A=\left(\begin{array}[c]{c}
                 \cosh(\theta_\xi/2)\\
                  -i\sinh(\theta_\xi/2)
      \end{array}\right)
\end{equation}

and 
\begin{equation}
B=\left(\begin{array}[c]{c}
                 i\sinh(\theta_\xi/2)\\
                  \cosh(\theta_\xi/2)
      \end{array}\right)
\end{equation}
abd subsequently the above Eq.~(\ref{matrix1}) can be easily reduced to
\begin{eqnarray}
 &&\la n,k_y|\hat{\mathcal{V}}_x|n',k'_y\ra\nonumber\\&&=
 -i\frac{v_c}{2\gamma_{\xi}}[\lambda\delta_{n-1,n'}+\lambda'\delta_{n,n'-1}]\delta_{k_y,k_y'}.
\end{eqnarray}
Similarly, the another matrix elements
\begin{eqnarray}\label{matrix2}
 &&\la n',k'_y|\hat{\mathcal{V}}_y|n,k_y\ra=\la n'|v_{e}^{\xi}\mathbb{1}+\xi v_{c}\hat{\sigma}_{y}|n\ra\delta_{k_y,k_y'}\nonumber\\
 &=&-\frac{1}{2\gamma_{\xi}}[v_c\cosh(\theta_{\xi})-v_{e}^{\xi}\sinh(\theta_{\xi})]\nonumber\\&\times&
 [\lambda'\delta_{n',n-1}+\lambda\delta_{n'-1,n}]
\end{eqnarray}
Now, we shall use the relation $\cosh(\theta_\xi)=\gamma_\xi$ and $\sinh(\theta_\xi)=\beta_\xi\gamma_\xi$ to obtain
\begin{eqnarray}
 \la n',k'_y|\hat{\mathcal{V}}_y|n,k_y\ra&=&-\frac{v_c}{2}(1-\beta_\xi^2)\nonumber\\&&
 [\lambda'\delta_{n',n-1}+\lambda\delta_{n'-1,n}]\delta_{k_y,k_y'}.
\end{eqnarray}
\section{}
To calculate the scattering rate between two states, the scattering matrix can be expressed as
\begin{eqnarray}
F_{\zeta,\zeta'}&=&\la \zeta|e^{i{\bf q.r}}|\zeta'\ra= \la n, k_y|e^{i{\bf q.r}}|n',k'_y\ra\nonumber\\
&\simeq& e^{-\eta/2}R_{n,n'}(\eta)\delta_{k'_y,k_y-q_y}e^{-i\Theta-\frac{\eta}{2}}
\end{eqnarray}
 with $\Theta=l_cq_x(-k_y+q_y/2)$. Here, we have also used the following standard integral results as:
 for $n'\ge n$
 \begin{equation}
  R_{n,n'}=\left[\frac{2^nn!}{2^{n'}n'!}\right]^{1/2}\upsilon^{n'-n}L_{n}^{n'-n}(\eta)e^{-\eta/2}
 \end{equation}
and for $n'\le n$
 \begin{equation}
  R_{n,n'}=\left[\frac{2^nn'!}{2^{n'}n!}\right]^{1/2}(-\upsilon^{*})^{n-n'}L_{n'}^{n-n}(\eta)e^{-\eta/2},
 \end{equation}
 where $\upsilon=l_c(q_x+iq_y)/2$ and $L_{n}^{n'}(\eta)$ is the Laguerre polymonial.
 More details about these matrix elements evaluation can be found in
 Refs.~[\onlinecite{PhysRevB.83.075427,PhysRevB.46.4667}].
\end{appendix}
\begin{acknowledgements}
Author acknowledges Alestin Mawrie for useful comments.
\end{acknowledgements}
\bibliography{bibfile}{}
\end{document}